\documentclass[aps,prd,preprint,superscriptaddress,preprintnumbers,eqsecnum,nofootinbib,nobibnotes,showpacs]{revtex4}
\usepackage{amsfonts,amssymb,amsmath,bm}
\usepackage{graphicx}
\usepackage{pstricks,color}

\newcommand{\be}{\begin{equation}}
\newcommand{\bea}{\begin{eqnarray}}
\newcommand{\ee}{\end{equation}}
\newcommand{\eea}{\end{eqnarray}}
\newcommand{\bpi}{\begin{picture}}
\newcommand{\bce}{\begin{center}}
\newcommand{\epi}{\end{picture}}
\newcommand{\ece}{\end{center}}

\newcommand{\D}{\displaystyle}
\def\chic#1{{\scriptscriptstyle #1}}

\newcommand{\qe}{{q}}

\def\gb{\bm{\Gamma}}

\def\l{\bm{\mathrm L}}
\def\g{\widetilde\gb}

\def\ie{{\it i.e.}, }

\begin{document}

\title{Nonperturbative gluon and ghost propagators for $d=3$ Yang-Mills}
\date{29.01.2010}

\author{A.~C. Aguilar}
\affiliation{Federal University of ABC, CCNH, 
Rua Santa Ad\'elia 166,  CEP 09210-170, Santo Andr\'e, Brazil.}
\author{D.~Binosi}
\affiliation{European Centre for Theoretical Studies in Nuclear
  Physics and Related Areas (ECT*), Villa Tambosi, Strada delle
  Tabarelle 286, I-38050 Villazzano (TN), Italy.}
\author{J. Papavassiliou}
\affiliation{Department of Theoretical Physics and IFIC, 
University of Valencia-CSIC,
E-46100, Valencia, Spain.}

\begin{abstract}

We study a manifestly gauge invariant set of 
Schwinger-Dyson  equations to determine the 
nonperturbative dynamics of the  gluon and ghost propagators in $d=3$ Yang-Mills.
The use of the well-known Schwinger mechanism, in the Landau gauge, leads to 
the dynamical generation of a mass for the gauge boson (gluon in $d=3$), 
which, in turn, gives rise to an infrared finite gluon propagator and ghost dressing function. 
The propagators obtained from the numerical solution of these nonperturbative 
equations are in very good agreement with the results of $SU(2)$ lattice simulations.

\end{abstract}

\pacs{
12.38.Lg, 
12.38.Aw,  
12.38.Gc   
}

\maketitle

\section{Introduction}

QCD in three space-time dimensions ($\rm{QCD}_3$ for short)
has received increasing attention in recent years, not only 
because it is the infinite-temperature limit of 
its four-dimensional counterpart ($\rm{QCD}_4$), 
but also because, at zero temperature,  
these two theories, despite a number of important differences, 
seem to share a variety of important nonpertubative features~\cite{Gross:1980br,Jackiw:1980kv,Appelquist:1981vg,Deser:1981wh,
Deser:1982vy,Nadkarni:1982kb,Cornwall:1984eu,Nadkarni:1988ti,Cornwall:1988ad,Kobes:1990dc,
Cornwall:1992cu,Buchmuller:1994qy,Jackiw:1995nf,Alexanian:1995rp,Cornwall:1995ac,Cornwall:1996jb,Karsch:1996aw,
Jackiw:1997jga,Buchmuller:1996pp,Cornwall:1997dc,Karabali:1998yq,Eberlein:1998yk,Heller:1997nqa,
Eberlein:1998mb,Cucchieri:2001tw,Nakamura:2003pu,Burgio:2009xp,Quandt:2010yq}
. 

$\rm{QCD}_3$ differs from $\rm{QCD}_4$ in several aspects.
For example, the fact that $\rm{QCD}_3$ lives in an odd-dimensional space 
allows the appearance of phenomena that are not possible in even-dimensional spaces, 
such as the parity violating gauge-boson masses from a Chern-Simons term~\cite{Jackiw:1980kv,Deser:1981wh,Deser:1982vy}. 
In addition, unlike $\rm{QCD}_4$, there is no linearly rising potential 
for the quarks. Moreover, given that in $d=3$ the 
square of the coupling constant has dimensions of mass, 
$\rm{QCD}_3$ is super-renormalizable, having a trivial renormalization group.  
Finally, there are no finite-action classical solitons in $\rm{QCD}_3$ (\ie no instantons)
[see~\cite{Cornwall:1988ad} for a brief review].     

On the other hand, both theories confine, 
display area laws for Wilson loops in the fundamental representation, 
and develop nonperturbative vacuum condensates, such as 
$\mathrm{Tr}\langle \mathcal{G}_{ij}^2\rangle $;  in fact, in $d=3$ one can 
actually prove~\cite{Cornwall:1992cu} the existence of a $\mathrm{Tr}\langle \mathcal{G}_{ij}^2\rangle $
condensate, associated with the minimum of the zero-momentum effective action, 
simply on the hypothesis that the full
theory possesses a unique mass scale (that of the gauge coupling).  
In addition, and more importantly for the purposes of the present work, 
both theories appear to cure their infrared (IR) instabilities through the dynamical generation of a 
gauge boson (gluon) mass, usually refereed to also as ``magnetic'' mass, 
 without affecting    the   local gauge invariance,    which   remains
intact~\cite{Cornwall:1982zr}.    
The nonperturbative dynamics  that gives rise to the generation of such 
a mass is rather complex, and 
can be ultimately traced back to a subtle realization 
of the Schwinger mechanism~\cite{Schwinger:1962tn,Schwinger:1962tp,Jackiw:1973tr,Cornwall:1973ts,Eichten:1974et,
Jackiw:1973ha,Farhi:1982vt,Frampton:2008zz}.

The  gluon  mass generation  manifests  itself  at  the level  of  the
fundamental Green's  functions of the  theory in a very  distinct way,
giving rise to an IR behavior that would be difficult to explain
otherwise.  Specifically,  in the Landau gauge, both  in $d=3,4$, 
the gluon  propagator and the  ghost dressing function reach  a finite
value  in the deep  IR. 
However, the gluon
propagator of $\rm{QCD}_3$ displays  a local maximum at relatively low
momenta, before reaching a  finite value at $q=0$. 
This characteristic
behavior is  qualitatively different to what happens  in $d=4$, where
the gluon  propagator is a monotonic  function of the  momentum in the
entire range between the IR and UV fixed points~\cite{Aguilar:2008xm}.

It should also be mentioned 
that a qualitatively similar situation emerges within the 
``refined'' Gribov-Zwanziger formalism~\cite{Gribov:1977wm,Zwanziger:1993dh}, presented in~\cite{Dudal:2008sp}. 
In this latter framework the gluon mass is obtained   
through the addition of appropriate condensates 
to the original Gribov-Zwanziger action.

Even though several aspects of $\rm{QCD}_3$
have been studied in a variety of works, the 
recent theoretical developments associated with the pinch technique (PT),
together with the high-quality lattice results produced, 
motivate the detailed study of the entire shape of the 
gluon and ghost propagators in $d=3$. 
Specifically, given that the gluon mass generation 
is a purely nonperturbative effect, in the continuum 
it has to be addressed within the framework  
of the Schwinger-Dyson equations (SDE). 
 These complicated dynamical equations are best studied in
a  gauge-invariant framework based on the  pinch technique  
(PT)~\cite{Cornwall:1982zr,Cornwall:1989gv,Binosi:2002ft,Binosi:2009qm,Nair:2005iw}, and its profound  
correspondence with the background field method (BFM)~\cite{Abbott:1980hw}.
As has been explained in detail in the recent literature~\cite{Aguilar:2006gr,Binosi:2007pi}, 
this latter formalism  
allows for a  gauge-invariant truncation of the SD series, 
in the sense that it preserves manifestly and at every step 
the transversality of the gluon self-energy.

In the present work we study the dynamics of the gluon and ghost propagators 
of pure Yang-Mills in $d=3$, using the SDEs  
of the PT-BFM formalism in the Landau gauge. 
Even though our results are 
valid for every gauge group, we will eventually 
focus on the group $SU(2)$, in order to make contact with available 
lattice simulations~\cite{Cucchieri:2003di}. The crucial ingredient in this analysis, 
which accounts for the type of solutions obtained, is the 
gauge-invariant introduction of a gluon mass. The way gauge invariance is 
maintained is through the inclusion 
of Nambu-Goldstone-like (composite) massless excitations
into the non-perturbative three-gluon vertex~\cite{Cornwall:1982zr}. As a result, 
the fundamental Ward identities of the theory, which encode the 
underlying gauge symmetry, remain intact. 
The results obtained from our SDE analysis, 
presented in section 4, compare rather well 
with the available lattice data [see in particular Figs~\ref{bestfit-figure} and~\ref{gluon-ghost-figure}].

In addition, as a necessary intermediate step, 
we calculate an auxiliary function, denoted by $G(q)$, 
which plays an instrumental role in the PT-BFM framework (see next section). 
Interestingly enough, and in the Landau gauge {\it only}, $G(q)$ coincides with the 
so-called Kugo-Ojima (KO) function; this latter function,
and in particular its value in the deep IR, is intimately
connected to the corresponding and well-known confinement criterion~\cite{Kugo:1979gm}. 
 
The article is organized as follows.
In Section II we briefly review the salient features of the SDEs within the 
PT-BFM framework. Section III contains a general discussion of the 
main conceptual issues related with the dynamical mass generation
through the Schwinger mechanism. Particular attention is paid to the 
specific form of the three-gluon vertex that must be employed in order to maintain 
gauge invariance, in the form of the Ward identities. In addition, we 
give a qualitative discussion of some of the main features expected 
for the gluon propagator in the presence of a gluon mass.  
Section IV contains the main results of this work. 
After setting up the corresponding SDE for the gluon propagator and the 
auxiliary function $G(q)$, we give explicit closed expressions for 
the latter quantities. The two available free parameters appearing in the 
expression for the gluon propagator, namely
the gauge coupling $g$ and the mass $m$
are then varied, in order to 
obtain the best possible agreement with the lattice data.  
The ghost dressing function is also obtained from the self-consistent 
solution of the corresponding SDE; it too shows a good agreement with the lattice. 
Finally, in Section V we present our conclusions.

\section{The PT-BFM framework}

In this section we remind the reader the basic characteristics of the 
SD framework that is based on the 
PT-BFM formalism; for an extended review of the subject see~\cite{Binosi:2009qm}. 

We start by introducing the necessary notation. The gluon propagator $\Delta_{\mu\nu}(q)$ 
in the covariant gauges assumes the form 
\be
\Delta_{\mu\nu}(q)= -i\left[ {P}_{\mu\nu}(q)\Delta(q) +\xi\frac{\D q_\mu
q_\nu}{\D q^4}\right],
\label{prop_cov}
\ee
where $\xi$ denotes the gauge-fixing parameter, 
\mbox{$P_{\mu\nu}(q)= g_{\mu\nu} - q_\mu q_\nu /q^2$}
is the usual transverse projector, and $\Delta^{-1}(q) = q^2 + i \Pi(q)$, 
with  $\Pi_{\mu\nu}(q)=P_{\mu\nu}(q) \Pi(q)$ the gluon self-energy. 
We also define the dimensionless 
vacuum-polarization ${\bf \Pi}(q)$, as $\Pi(q)= q^2 {\bf \Pi}(q)$. 
In addition, the full ghost propagator, $D(p)$,
 and its dressing function, $F(p)$ , are related by $D(p)= iF(p)/p^2$.

\begin{figure}[!t]
\begin{center}
\includegraphics[width=16cm]{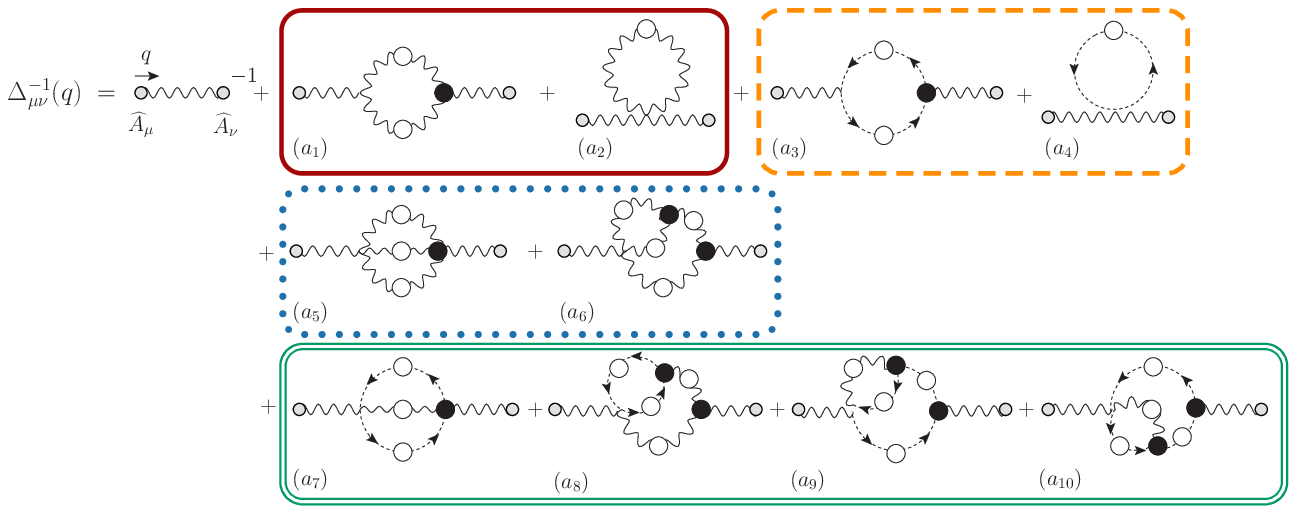}
\end{center}
\vspace{-1.0cm}
\caption{\label{sde_bfm}The full SDE for the gluon self-energy in the PT-BFM framework. By virtue of the 
special Abeliean-like Ward identities satisfied by the various fully dressed vertices, 
the contributions of each block are individually transverse.}
\end{figure}

The truncation scheme for the SDEs of Yang-Mills theories based on the PT 
respects gauge  invariance (\ie the transversality of the gluon self-energy)  
at every level  of  the ``dressed-loop''
expansion.  This becomes  possible  due to  the drastic  modifications
implemented  to  the building  blocks  of  the  SD series, \ie  the
off-shell  Green's  functions themselves,  following 
the  general methodology of the PT~\cite{Cornwall:1982zr,Cornwall:1989gv,Nair:2005iw}.
The PT  is a well-defined  algorithm that exploits  systematically the
BRST symmetry in order to construct new Green's functions endowed with
very  special  properties; in particular, the crucial property of gauge invariance, for 
they satisfy  Abelian Ward identities instead  of the usual Slavnov-Taylor identities
The PT may be used to rearrange systematically the entire SD series~\cite{Aguilar:2006gr}. In the case of  
the gluon self-energy it gives rise to a new SDE, shown schematically in Fig.~\ref{sde_bfm}. 

Note that the quantity that appears on the lhs of Fig.~\ref{sde_bfm} is {\it not} 
the conventional self-energy $\Pi_{\mu\nu}$, but rather 
the PT-BFM self-energy, denoted by $\widehat{\Pi}_{\mu\nu}$.
The graphs appearing on the rhs  
contain the conventional self-energy $\Pi_{\mu\nu}$ as before, 
but are composed out of two types of vertices:
\begin{enumerate}
\item[(i)] The conventional vertices, where all incoming fields are 
quantum fields, \ie they carry the 
virtual loop momenta; these vertices are all ``internal'', \ie the external gluons cannot 
be one of their legs, and will be generally denoted by ${\gb}$.
\item[(ii)] A new set of vertices, with one of their legs being the 
external gluon, carrying physical momentum $q$; 
these new vertices, to be generally denoted by ${\g}$, correspond precisely to the Feynman rules of the BFM~\cite{Abbott:1980hw}, \ie
it is as if the external gluon had been converted dynamically into a background gluon. 
\end{enumerate}

As a result,
the full vertices ${\g}_{\alpha\mu\nu}^{amn}(q,k_1,k_2)$, ${\g}_{\alpha}^{anm}(q,k_1,k_2)$,
${\g}_{\alpha\mu\nu\rho}^{amnr}(q,k_1,k_2,k_3)$, and ${\g}_{\alpha\mu}^{amnr}(q,k_1,k_2,k_3)$
appearing on the rhs of the SDE shown in Fig.~\ref{sde_bfm} satisfy the simple Ward identities 
\bea
q^{\alpha}{\g}_{\alpha\mu\nu}^{amn} &=&
gf^{amn}
\left[\Delta^{-1}_{\mu\nu}(k_1)
- \Delta^{-1}_{\mu\nu}(k_2)\right],
\nonumber\\
q^{\alpha}{\g}_{\alpha}^{anm} &=&  igf^{amn}
\left[D^{-1}(k_1)- D^{-1}(k_2)\right],
\nonumber\\ 
q^{\alpha}{\g}_{\alpha\mu\nu\rho}^{amnr} &=&
g f^{adr} {\gb}_{\nu\rho\mu}^{drm}(q+k_2,k_3,k_1) + {\rm c.p.},
\nonumber\\
q^{\alpha} {\g}_{\alpha\mu}^{amnr} &=&
g f^{aem} {\gb}_{\mu}^{enr}(q+k_1,k_2,k_3) + {\rm c.p.}, 
\label{fourWI}
\eea
where ``cp'' stands for ``cyclic permutations''. 
Using these identities, it is straightforward to show 
that the crucial transversality condition $q^{\mu} \widehat{\Pi}_{\mu\nu}(q) = 0$ 
is enforced ``block-wise''~\cite{Aguilar:2006gr}, \ie  
\bea
q^{\mu} [(a_1)+(a_2)]_{\mu\nu} &=& 0\,,
\nonumber\\
q^{\mu} [(a_3)+(a_4)]_{\mu\nu} &=& 0\,,
\nonumber\\
q^{\mu} [(a_5)+(a_6)]_{\mu\nu} &=& 0\,,
\nonumber\\
q^{\mu} [(a_7)+(a_8)+(a_9)+(a_{10})]_{\mu\nu} &=& 0\,,
\label{tranbl}
\eea
which allow for a self-consistent truncation of the full gluon SDE given in Fig.~\ref{sde_bfm}.  

\begin{figure}[!t]
\begin{center}
\includegraphics[scale=0.75]{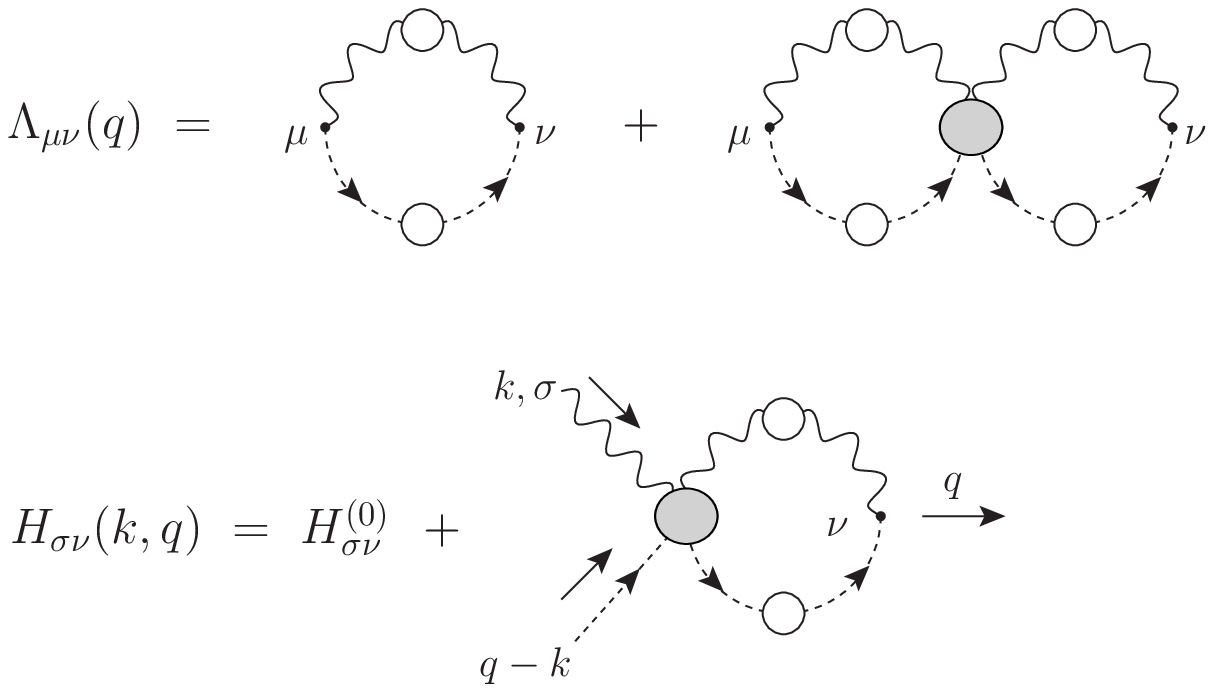}
\vspace{-0.5cm}
\caption{\label{Lambda-H} Diagrammatic representation of the functions $\Lambda$ and $H$.}
\end{center}
\end{figure}

Quite interestingly, the conventional $\Delta(q)$ and its PT-BFM counterpart $\widehat{\Delta}(q)$
the two quantities are 
connected by the following {\it background-quantum} identity~\cite{Grassi:1999tp}
\be
\Delta(q) = 
\left[1+G(q)\right]^2 \widehat{\Delta}(q), 
\label{bqi2}
\ee
where the function $G(q)$
is the $g_{\mu\nu}$ component
of the auxiliary two-point function $\Lambda_{\mu \nu}(q)$, 
defined as
\bea
\Lambda_{\mu \nu}(q) &=&-i g^2C_A
\int_k H^{(0)}_{\mu\rho}
D(k+q)\Delta^{\rho\sigma}(k)\, H_{\sigma\nu}(k,q),
\nonumber \\
&=& g_{\mu\nu} G(q) + \frac{q_{\mu}q_{\nu}}{q^2} L(q),
\label{LDec}
\eea
where $C_{\rm {A}}$ the Casimir eigenvalue of the adjoint representation
[$C_{\rm {A}}=N$ for $SU(N)$] and 
 \mbox{$\int_{k}\equiv\mu^{2\varepsilon}(2\pi)^{-d}\int\!d^d k$}, with $d$ the dimension of space-time.
The function $H_{\sigma\nu}$ is given diagrammatically in Fig.~\ref{Lambda-H}.
Note that it is related to the full gluon-ghost vertex by 
\mbox{$q^{\sigma} H_{\sigma\nu}(p,r,q) = -i{\gb}_{\nu}(p,r,q)$}; 
at tree-level, $H_{\sigma\nu}^{(0)} = ig_{\sigma\nu}$.

The identity (\ref{bqi2}) allows to express the SDE of  Fig.~\ref{sde_bfm} 
as an integral equation involving only $\Delta(q)$, namely  
\be
\Delta^{-1}(q)P_{\mu\nu}(q) = 
\frac{q^2 P_{\mu\nu}(q) + i \sum_{i=1}^{10}(a_i)_{\mu\nu}}{[1+G(q)]^2}. 
\label{SDgl}
\ee

\begin{figure}[!t]
\begin{center}
\includegraphics[scale=0.8]{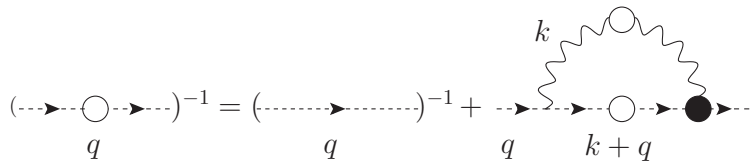}
\vspace{-0.2cm}
\caption{\label{ghostSDE}The SDE satisfied by the ghost propagator.}
\end{center}
\end{figure}

Finally, as shown in Fig.~\ref{ghostSDE}, the ghost SDE is the same as in the conventional formulation, namely 
\be
iD^{-1}(q) = q^2 + i g^2 C_A  \int_k
\Gamma^{\mu}\Delta_{\mu\nu}(k)\gb^{\nu}(q,k) D(q+k),
\label{SDgh}
\ee
where $\Gamma_{\mu}$ is the standard (asymmetric) gluon-ghost vertex at tree-level,
and $\gb^{\mu}$ it fully-dressed counterpart. 

\section{Mass generation in $d=3$ Yang-Mills}

It is well-known that, just as happens at $d=4$,  the Yang-Mills dynamics in $d=3$ 
generates an effective gauge-boson mass, which cures all IR instabilities. 
The underlying mechanism that leads to the generation of such a dynamical mass, 
both in $d=3,4$, is the Schwinger mechanism, the  
only known procedure for obtaining massive gauge bosons while 
maintaining the gauge-symmetry intact. 

As Schwinger pointed out long time ago~\cite{Schwinger:1962tn},     
the gauge invariance of a vector field does not necessarily 
imply zero mass for the associated particle, if the 
current vector coupling is sufficiently strong. 
According to Schwinger's fundamental observation, 
if ${\bf \Pi}(q)$ 
acquires a pole at zero momentum transfer, then the 
 vector meson becomes massive, even if the gauge symmetry 
forbids a mass at the level of the fundamental Lagrangian.
Indeed, it is clear that if the vacuum polarization ${\bf \Pi}(q)$ 
has a pole at  $q^2=0$ with positive 
residue $\mu^2$, \ie ${\bf \Pi}(q) = \mu^2/q^2$, then (in Euclidean space)
$\Delta^{-1}(q) = q^2 + \mu^2$.
Thus, the vector meson 
becomes massive, $\Delta^{-1}(0) = \mu^2$, 
even though it is massless in the absence of interactions ($g=0$). 
There is {\it no} physical principle which would preclude ${\bf \Pi}(q)$ from 
acquiring such a pole, even in the absence of elementary scalar fields. 
In a {\it strongly-coupled} theory, like non-perturbative Yang-Mills in $d=3,4$, 
this may happen for purely dynamical reasons,
since strong binding may generate zero-mass bound-state excitations~\cite{Jackiw:1973tr}.
The latter  act  {\it  like}
dynamical Nambu-Goldstone bosons, in the sense that they are massless,
composite,  and {\it longitudinally   coupled};  but, at  the same  time, they
differ  from  Nambu-Goldstone  bosons   as  far  as  their  origin  is
concerned: they  do {\it not} originate from  the spontaneous breaking
of  any global symmetry~\cite{Cornwall:1982zr}.
In what follows we will assume that theory can  
indeed generate the required bound-state poles;  
the demonstration of the existence of a bound state, and in particular of a zero-mass bound state, 
is a difficult dynamical problem, usually  
studied by means of integral equations known as Bethe-Salpeter equations
(see, e.g.,\cite{Poggio:1974qs}). Note also that the 
generation of a dynamical mass (both in $d=3,4$) 
requires (and, correspondingly, gives rise to), the formation of a gluon condensate. 

\begin{figure}[!t]
\begin{center}
\includegraphics[scale=.75]{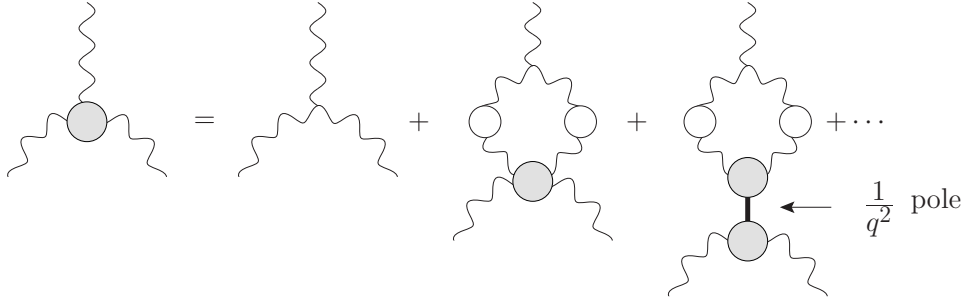}
\vspace{-0.3cm}
\caption{\label{bound_new}Vertex with  non-perturbative massless excitations triggering the Schwinger mechanism.}
\end{center}
\end{figure}

The Schwinger mechanism
is incorporated into the SDE of the gluon propagator essentially 
through the form of the fully-dressed, nonperturbative three-gluon vertex. 
In fact, since the generation of the mass does not interfere 
with the gauge symmetry, 
which remains intact, the three-gluon 
vertex must satisfy the same Ward identity as in the massless case [{\it viz.} Eq.(\ref{fourWI})], 
but now with massive, as opposed to 
massless, gluon propagators on its rhs.   
The way this crucial requirement is enforced is precisely through the incorporation into the 
three-gluon vertex of 
the Nambu-Goldstone (composite) massless excitations mentioned above. 
To see how this works with a simple example, 
let us consider the standard tree-level vertex 
\be
\Gamma_{\mu\alpha\beta}(q,p,r) = (q-p)_{\beta}g_{\mu\alpha} 
+ (p-r)_{\mu}g_{\alpha\beta} + (r-q)_{\alpha}g_{\mu\beta}\,,
\label{stanvert}
\ee
which satisfies the simple Ward identity 
\be
q^{\mu}\Gamma_{\mu\alpha\beta}(q,p,r) = 
P_{\alpha\beta}(r) \Delta_0^{-1}(r) - P_{\alpha\beta}(p) \Delta_0^{-1}(p)
\label{ewi}
\ee
where $\Delta_0^{-1}(q) = q^2$ is the inverse of the tree-level propagator. 
After the dynamical mass generation, the inverse gluon propagator becomes, roughly speaking,  
\be
\Delta_m^{-1}(q)= q^2 - m^2(q^2), 
\label{rg5}
\ee
and the new vertex, $\gb_{\mu\alpha\beta}^{m}(q,p,r)$ that replaces 
$\Gamma_{\mu\alpha\beta}(q,p,r)$ must still satisfy the Ward identity of (\ref{ewi}), but with 
 $\Delta_0^{-1} \to \Delta_m^{-1}$ on the rhs.
This is accomplished if   
\be
\gb^{m}_{\mu\alpha\beta}(q,p,r) = \Gamma_{\mu\alpha\beta}(q,p,r) + 
V_{\mu\alpha\beta}(q,p,r),
\label{verv}
\ee
where $V_{\mu\alpha\beta}(q,p,r)$ contains the massless poles. A standard Ansatz for 
 $V_{\mu\alpha\beta}(q,p,r)$ is~\cite{Cornwall:1984eu} 
\be
V_{\mu\alpha\beta}(q,p,r) =
 m^2(r)\frac{q_{\mu} p_{\alpha}(q-p)_{\rho}}{2q^{2}p^{2}}\,
P^{\rho}_{\beta}(r)  - 
\left[m^2(p)-m^2(q)\right]\frac{r_{\beta}}{r^{2}}\,P^{\mu}_{\rho}(q)\,P^{\rho}_{\alpha}(p)  
\nonumber\\
+{\rm c.p.}\,, 
\label{Ver1}
\ee
It is easy to check that 
\be
q^{\mu}V_{\mu\alpha\beta}(q,p,r) = P_{\alpha\beta}(p) m^2(p) -  P_{\alpha\beta}(r) m^2(r)\,,
\label{wiv}
\ee
and cyclic permutations.  
Therefore, one has
\be
q^{\mu}\Gamma^{m}_{\mu\alpha\beta}(q,p,r) = 
P_{\alpha\beta}(r) \Delta_m^{-1}(r) - P_{\alpha\beta}(p) \Delta_m^{-1}(p)\,,
\label{ewim}
\ee
as announced. 
Note that for constant masses [$m(q)=m(p)=m(r)=m$] 
the vertex of  (\ref{Ver1})  reduces to 
\be
V_{\mu\alpha\beta}(q,p,r) = 
\frac{m^2}{2}
\left[\frac{q_{\mu} p_{\alpha}(q-p)_{\rho}}{q^{2} p^{2}} \,P^{\rho}_{\beta}(r)
+\frac{p_{\alpha} r_{\beta}(p-r)_{\rho}}{p^{2} r^{2}} \,P^{\rho}_{\mu}(q)
+ \frac{r_{\beta} q_{\mu}(r-q)_{\rho}}{r^{2} q^{2}} \,P^{\rho}_{\alpha}(p)\right].
\label{Vers}
\ee 
Even though the precise implementation 
at the level of the complicated integral equations 
is rather subtle, 
the final upshot of introducing a vertex 
such as $\gb^{m}$ (or more sophisticated versions of it)
into the SDE for the gluon self-energy 
is that one finally obtains, gauge-invariantly, a non-vanishing $\widehat\Delta^{-1}(0)$ 
and $\Delta^{-1}(0)$.
Qualitatively speaking, 
in Euclidean space and $d$ space-time dimensions, 
the (background) gluon propagator is given by 
\be
\widehat\Delta^{-1} (q) = q^2  + \widehat\Pi(q) + \widehat\Delta^{-1}(0), 
\ee
where $\widehat\Pi(q)$ has the general form
\be
\widehat\Pi(q) = c_1 g^2 \int_k \Delta(k)\Delta(k+q) K_1(q,k) + 
c_2 g^2 \int_k D(k)D(k+q) K_2(q,k).
\label{Jd}
\ee
The functions $K_1(q,k)$ and $K_2(q,k)$
are SD kernels, whose closed form 
depends, among other things, on the dimensionality of space-time, the details of the 
vertices employed, and the gauge chosen, as do, in general,  the constants 
$c_1$ and $c_2$.   
Setting $\widehat\Delta^{-1}(0) = m^2$, one then obtains 
\be
\widehat\Delta^{-1} (q) = q^2 + m^2 + \widehat\Pi(q). 
\ee
To obtain the perturbative (one-loop) expression for $\widehat\Delta(q)$ 
one must substitute in the integral on the rhs of (\ref{Jd}) the 
tree-level values for $\Delta$, $D$, $K_1$ and $K_2$, which is a 
good approximation for large values of the physical momentum $q$. 
However, for low values of $q$, 
one must solve the integral equation, 
which, under suitable assumptions, will furnish massive (IR finite)
solutions for $\widehat\Delta(q)$. 

An easy  way to qualitatively appreciate the effect of the mass on the solutions 
for $\widehat\Delta(q)$ 
is to substitute $\Delta \to \Delta_m$ in the first integral on the rhs of (\ref{Jd}), 
assuming for simplicity a constant mass $m$, 
and use tree-level expressions for all other terms. This will furnish an approximate 
expression for $\widehat\Pi(q)$, to be denoted by  $\widehat\Pi_{m}(q)$, and the 
resulting $\widehat\Delta^{-1}(q)$ will 
read  
\be
\widehat\Delta^{-1}(q) = q^2 + m^2 + \widehat\Pi_{m}(q).
\label{dmp}
\ee  

In $d=4$ the corresponding $\widehat\Pi_m(q)$
will have the form 
\be
\widehat\Pi_m^{(4)}(q)  = bg^2 q^2 \int_0^1 dx \ln [q^2 x(1-x) + m^2].
\label{J4m}
\ee
For $m\to 0 $,  or $q^2\ll m^2$, 
one recovers the usual one-loop logarithm $bg^2 \ln (q^2)$, with 
 $b$ being the first coefficient of the QCD one-loop $\beta$-function, 
$b = 11 C_{\rm {A}}/ 48\pi^2 $. As explained in the literature, the presence 
of the mass inside the logarithm tames the Landau pole, and gives eventually 
rise to a IR finite value for the QCD effective charge

Similarly, in $d=3$ we have [see the integral $R_1$ in Eq.~(\ref{integrals})]
\be
\widehat\Pi_m^{(3)}(q)  = -2 b_3 g^2 {\qe} \arctan\left(\frac{\qe}{m}\right),
\label{J3m}
\ee
which in the limit $m \to 0$ assumes the one loop perturbative form [see also the integral $I_1$ in Eq.~(\ref{integrals})] 
\be
\widehat\Pi^{(3)}_{\rm pert}(q) = -{\pi b_3 g^2}{\qe}. 
\label{J3pert}
\ee
In this case however, and unlike in $d=4$, $b_3$ is a numerical coefficient that depends explicitly on the 
value of the gauge parameter chosen; in the Feynman gauge, $b_3=15 C_{\rm {A}}/32 \pi $ 
(we will return to this point in the next section). 

Let us now briefly compare the versions of the gluon propagator obtained by substituting 
$\Pi^{(3)}_{\rm pert}(q)$ or $\Pi_m^{(3)}(q)$ into (\ref{dmp}). 
For the perturbative case we have 
\be
\widehat\Delta_{\rm pert}(q) = \frac{1}{q^2 - \pi b_3 g^2 q}.
\label{dpert}
\ee  
There two points to notice: (i) $\widehat\Delta_{\rm pert}(q)$ has a Landau pole at   
$\overline{q}= \pi b_3 g^2$, and (ii) it displays a maximum value at  $q^{*}= \overline{q}/2$.    
On the other hand, the gluon propagator corresponding to $\widehat\Pi_m^{(3)}(q)$ becomes
\be
\widehat\Delta(q) = \frac{1}{q^2 + m^2 -  2 b_3 g^2 q \arctan\left(\frac{\qe}{m}\right)}.
\label{dpertm}
\ee 
It is clear that the presence of the mass regulates the denominator for all values of $q$, provided 
that it exceeds a certain critical value (in units of $g^2$). In addition, 
$\widehat\Delta(q)$ may or may not display a maximum, depending on the ratio $g^2/m$; 
in general, its position is displaced with respect to $q^{*}$.

\section{Results and Comparison with the lattice.}

In order to make contact with the $d=3$ lattice results of~\cite{Cucchieri:2003di}, we must next 
determine the form of the relevant SDEs  in the Landau gauge ($\xi =0$). 
The three quantities of interest are: 
\begin{itemize}
\item[(i)] The gluon propagator, $\Delta(q)$ given in (\ref{SDgl}); 
\item[(ii)] The Kugo-Ojima function $G(q)$, given in (\ref{LDec}), 
which connects the conventional and background gluon propagators;
\item[(iii)]  The ghost propagator,  given in (\ref{SDgh}), and in particular its dressing function, $F(q)$.  
\end{itemize}

\subsection{Calculating the gluon propagator(s) and the KO function}

In the ``one-loop dressed'' approximation, the PT-BFM gluon self-energy is given by   
the following (gauge-invariant) subset of diagrams:
\be
\widehat\Pi_{\mu\nu}(q) = [(a_1)+(a_2)+(a_3)+(a_4)]_{\mu\nu}.
\label{old}
\ee
When evaluating  the diagrams $(a_i)$  one should use the  BFM Feynman
rules~\cite{Abbott:1980hw},  noticing in particular  that the
bare  three- and  four-gluon  vertices depend  explicitly on  $1/\xi$,
the coupling of  the ghost to  a background gluon  is {\it
symmetric} in  the ghost  momenta,  and finally  that there is  a four-field
coupling between two  background gluons and two ghosts. 

As explained in \cite{Aguilar:2008xm}, the limit $\xi\to 0$ 
of the diagrams $(a_1)$ and $(a_2)$ must be taken with care,    
due to the terms proportional to  $1/\xi$ coming from the tree-level vertices.
Introducing $\Delta^\mathrm{t}_{\mu\nu}(q)=P_{\mu\nu}(q)\Delta(q)$, 
one obtains 
\bea
[(a_1)+(a_2)]_{\mu\nu} &=& g^2C_A \Bigg\{\frac{1}{2}\int_k
\Gamma_\mu^{\alpha\beta}\Delta_{\alpha\rho}^\mathrm{t}(k)\Delta_{\beta\sigma}^\mathrm{t}(k+q)\l_\nu^{\rho\sigma}
-\frac{4}{3}g_{\mu\nu}\int_k \Delta(k) \nonumber \\
&+&\int_k \!\!\Delta_{\alpha\mu}^\mathrm{t}(k) 
\frac{(k+q)_{\beta}}{(k+q)^2}[\Gamma+ {\l}]_\nu^{\alpha\beta}
+\int_k \frac{k_{\mu}(k+q)_{\nu}}{k^2 (k+q)^2}\Bigg\}.
\label{contr}
\eea
The vertex $\l_{\mu\alpha\beta}$ is the fully-dressed counterpart of $\Gamma_{\mu\alpha\beta}$ 
(in the Landau gauge); it satisfies the Ward identity
\be
q^{\mu} \l_{\mu\alpha\beta} = P_{\alpha\beta}(k+q)\Delta^{-1}(k+q)
-P_{\alpha\beta}(k)\Delta^{-1}(k) . 
\ee
It is then easy to verify that the 
rhs of (\ref{contr}) vanishes when contracted by $q^{\mu}$, thus 
explicitly confirming the validity of the first equation in (\ref{tranbl}),  
for the special case of $\xi =0$. 

Similarly, 
\be
[(a_3)+(a_4)]_{\mu\nu} = - g^2 C_A \bigg[\int_k  \widetilde{\Gamma}_{\mu} D(k) D(k+q) {\g}_{\nu}\,
- 2 i g_{\mu\nu} \int_k D(k)\bigg]\,,
\label{allgr}
\ee
with $\widetilde{\Gamma}_{\mu}(q,p,r)\sim  (r-p)_{\mu}$. The 
vertex ${\g}_{\mu}$ satisfies the second Ward identity in (\ref{fourWI}), 
which leads immediately to the transversality of this block, \ie the second equation in (\ref{tranbl}). 

Finally, using tree-level values for the auxiliary function $H_{\sigma\nu}$ in (\ref{LDec}) and for 
the vertex $\gb^{\nu}$ in (\ref{SDgh}), 
we obtain for the Kugo-Ojima function
\be
G(q) = \frac{g^2 C_A}{2}\int_k \left[1+ \frac{(k \cdot q)^2}{k^2 q^2}\right]\Delta (k)  D(k+q),
\label{Gapp}
\ee
while for $L(q)$ one has
\be
L(q)= \frac{g^2 C_A}{2}\int_k \left[1-3 \frac{(k \cdot q)^2}{k^2 q^2}\right]\Delta (k)  D(k+q).
\label{Lapp}
\ee

The way we proceed is the following. 
Instead of actually solving the system of coupled integral equation, we will adopt an approximate 
procedure, which is operationally less complicated, and seems to capture rather well the underlying dynamics. 

Specifically, we will assume that the gluon propagator has the form given 
in (\ref{dmp}), and will 
determine the function $\Pi_m^{(3)}(q)$ by calculating the expressions given 
in (\ref{contr}) and (\ref{allgr})  
using inside the corresponding integrals $\Delta\to \Delta_m$ and $D\to D_0$. In order 
to maintain gauge invariance intact, we will set 
\be
\l_{\mu\alpha\beta}(q,p,r) =  \gb^{m}_{\mu\alpha\beta}(q,p,r)
\ee 
with $\gb_{\mu\alpha\beta}^{m}(q,p,r)$ given in  (\ref{verv}). The vertex  
$V_{\mu\alpha\beta}^{m}(q,p,r)$ entering into $\Gamma_{\mu\alpha\beta}^{m}(q,p,r)$ will be that 
of Eq.~(\ref{Vers}), \ie we will assume a constant mass $m$ throughout. 

From the final expressions appearing in the rest of the paper we will use Euclidean momenta. 
To that end we set  $q^2 = -q^2_{\chic E}$, with  $q^2_{\chic E} >0$ the positive square of a 
Euclidean four-vector, and $q_{\chic E} = \sqrt{q^2_{\chic E}}$.
The Euclidean propagator is defined as $\widehat\Delta_{\chic E} (q^2_{\chic E}) = - \widehat\Delta (-q^2_{\chic E})$.  
To avoid notational clutter, we will suppress the subscript ``E'' in what follows. 

The results of all our calculations will be expressed in terms of the following six basic integrals,  
\bea
R_0&=&\int_k\frac1{k^2-m^2}=\left(\frac{i}{4\pi}\right) m,\nonumber \\
R_1&=&\int_k\frac1{(k^2-m^2)[(k+q)^2-m^2]}= \left(\frac{i}{4\pi}\right)\frac1{\qe}\arctan\left(\frac{\qe}{2m}\right),\nonumber \\
I_1&=&\int_k\frac1{k^2(k+q)^2}=\left(\frac{i}{8}\right) \frac1{\qe},\nonumber \\
I_2&=&\int_k\frac1{(k^2-m^2)(k+q)^2}= \left(\frac{i}{4\pi}\right)\frac1{\qe}\arctan\left(\frac{\qe}{m}\right),\nonumber \\
I_3&=&\int_k\frac1{k^2(k^2-m^2)}= \left(\frac{i}{4\pi}\right)\frac1m,
\nonumber \\
I_4&=&\int_k\frac{q\cdot k}{(k^2-m^2)(k+q)^2}= \left(\frac{i}{8\pi}\right) \left[m+\frac{q^2-m^2}{q}\arctan\left(\frac{\qe}{m}\right)\right],
\label{integrals}
\eea
where the momentum $q$ appearing in the integrals on the lhs is Minkowskian, while the momentum $q$ appearing in the results on the rhs is Euclidean.

To facilitate the calculation, and  
since the transversality of $\widehat\Pi_{\mu\nu}(q)$ is guaranteed, one may set in (\ref{old}) 
$\widehat\Pi_{\mu\nu}(q) = P_{\mu\nu}(q) \widehat\Pi_m(q)$, and isolate 
$\widehat\Pi_m(q)$ by taking the trace of both sides, \ie 
\be
(d-1)\widehat\Pi_m(q) = [(a_1)+(a_2)+(a_3)+(a_4)]_{\mu}^{\mu}
\label{ptr}
\ee

For the different four contributions shown in Eq.~(\ref{contr}) we obtain the following results  
\bea
\frac{1}{2}\int_k
\Gamma_\mu^{\alpha\beta}\Delta_{\alpha\rho}^\mathrm{t}(k)\Delta_{\beta\sigma}^\mathrm{t}(k+q)\l^{\mu\rho\sigma}&=&9R_0+\left(\frac14\frac{\qe^6}{m^4}-2\frac{\qe^4}{m^2}-10\qe^2+8m^2\right)R_1+\frac14\frac{\qe^6}{m^4}I_1,\nonumber \\
&-&\left(\frac12\frac{\qe^6}{m^4}-2\frac{\qe^4}{m^2}-\frac{11}{2}\qe^2-3m^2\right)I_2-\left(\frac52\qe^2+4m^2\right)I_3,
\nonumber \\
\frac{4}{3}g^\mu_\mu\int_k \Delta(k)&=&4R_0,\nonumber \\
\int_k \!\!\Delta_{\alpha\mu}^\mathrm{t}(k) 
\frac{(k+q)_{\beta}}{(k+q)^2}[\Gamma+ {\l}]^{\mu\alpha\beta}&=&-\left(\frac12+\frac14\frac{m^2}{\qe^2}\right)R_0-\left(\frac12\frac{\qe^4}{m^2}+\frac14\qe^2\right)I_1\nonumber\\
&+&\left(\frac12\frac{\qe^2}{m^2}-\frac{11}4\qe^2-3m^2+\frac14\frac{m^4}{\qe^2}\right)I_2+\left(\frac12\qe^2+\frac{m^2}4
\right)I_3,\nonumber \\
\int_k \frac{k^{\mu}(k+q)_{\mu}}{k^2 (k+q)^2}&=&\frac12\qe^2 I_1.
\label{cona}
\eea
Next, let us turn to the diagrams $(a_3)$ and $(a_4)$ of Fig.~\ref{sde_bfm}, which contain a ghost loop. 
Since we will treat the ghost as a massless particle, 
the ``tadpole'' diagram $(a_4)$ vanishes identically in dimensional regularization;
from diagram $(a_3)$ we get  instead (after taking the trace)
\be
(a_3)_{\mu}^{\mu}=-g^2 C_A \int_k\frac{(2k+q)^\mu(2k+q)_\mu}{k^2(k+q)^2}=-g^2 C_A\qe^2 I_1.
\label{conb}
\ee

From the results above  
it is relatively straightforward to check, taking appropriate limits, that in the deep IR  
\be
\widehat{\Pi}(0) = -i\frac{g^2 C_A}{6\pi}m . 
\ee
Therefore, in order for the (Euclidean) 
$\widehat{\Delta}(0)^{-1}= m^2 - i \widehat{\Pi}(0) $ to be positive definite, 
$m$ and $g$ must satisfy the condition 
\be
\frac{m}{C_Ag^2} > \frac{1}{6\pi}.
\ee

In the opposite limit, namely for asymptotically large momenta, 
the addition of all terms given in (\ref{cona}) 
exposes a vast cancellation of all powers $q^n$, with $n>1$. 
After all such cancellations taking place, one is left with 
a linear contribution, given by
\be
\widehat{\Pi}(\qe)\stackrel{\qe\to\infty}{\longrightarrow}-i\frac{g^2 C_A}{32}\qe\left(15-\frac72\right).
\label{largeQ}
\ee
The reason for writing the numerical coefficient in front of the leading contribution  
as a deviation from 15 is the following. 
The expression (\ref{largeQ}) should coincide with the $d=3$ one-loop BFM self-energy
calculated in the Landau gauge. For any dimension  $d$ and any value of the gauge-fixing parameter $\xi_Q$, the latter   
reads~\cite{Binosi:2009qm} 
\bea
\widehat{\Pi}(q)  & = & \frac{g^2 C_A}{2}\left(\frac{7d-6}{d-1}\right)
q^2  \int_k \frac{1}{k^2 (k+q)^2}\nonumber \\
&-& g^2 C_A q^2(1-\xi_Q)
\left[\frac{1-\xi_Q}2q^2 {P}^{\mu\nu}(q) \int_k\!
\frac{k_{\mu} k_{\nu}}{k^4 (k+q)^4} 
+  \int_k\! \frac{2q\cdot k}{k^4 (k+q)^2}\right].
\label{genprop}
\eea
In the Feynman gauge of the BFM, $\xi_Q=1$, $\widehat{\Pi}(q)$ collapses to the PT answer for the 
gauge-independent gluon self-energy; specifically, for $d=3$,  
\be
\widehat{\Pi}(q)|_{\xi_Q=1} = -i\frac{g^2 C_A}{32}\qe (15). 
\ee
Away from $\xi_Q=1$ the terms in the second line of (\ref{genprop}) give additional 
contributions, which may be easily calculated using the basic results
\bea
\int_k\!\frac{k_{\mu} k_{\nu}}{k^4 (k+q)^4} &=&-\frac{i}{32}\frac1{\qe^3}g_{\mu\nu}+\cdots,\nonumber \\
\int_k\! \frac{q\cdot k}{k^4 (k+q)^2}&=&-\frac{1}{16\qe},
\eea
where the dots in the first integral indicate longitudinal parts. 
In particular, it is easy to verify that at $\xi_Q=0$ these additional terms 
account precisely for the term $-\frac72$ appearing in Eq.~(\ref{largeQ}). 

The above discussion reveals an important difference between the $d=3$ and $d=4$ cases.
Specifically, in $d=4$ the coefficient in front of the leading one-loop contribution
to $\widehat{\Pi}(\qe)$
is independent of the gauge-fixing parameter $\xi_Q$. 
This well-known BFM result
can be easily deduced from (\ref{genprop}), since 
both integrals proportional to $(1-\xi_Q)$ are UV finite, \ie they do not furnish 
logarithms. The coefficient in front of the logarithm is completely determined by the 
first integral, multiplied by the factor $\frac{g^2 C_A}{2}\left(\frac{7d-6}{d-1}\right)$, 
which, at $d=4$, reduces to  
\mbox{$16\pi^2b= (11/3)g^2 C_A$}, namely the first coefficient of the Yang-Mills $\beta$ function.  
As we have just demonstrated, 
things are different in $d=3$, where no renormalization is needed;    
the leading (linear) contribution depends explicitly on the value of $\xi_Q$.

Next, we determine an approximate expression for the function $G(q)$. To that end, we turn to (\ref{Gapp})
and substitute in the integral on the rhs, $\Delta \to \Delta_m$ and $D \to D_0 $. 
One has then
\bea
G(q)&=& \frac{i g^2 C_A}{2} \int_k\frac1{(k+q)^2 (k^2-m^2)}\left[1+\frac{(k\cdot q)^2}{k^2q^2}\right]
\nonumber\\
&=& \frac{ig^2C_A}8\left[-\frac2{q^2}I_4+5I_2-I_3+
\frac{q^2}{m^2}\left(I_2-I_1 \right)\right]
\label{Gint}
\eea
which gives
\be
G(q)= -\frac{g^2 C_A}{32\pi m}\left[\frac\pi2\frac{q}{m}+\frac{m^2}{q^2}-1+\left(6-\frac{m^2}{q^2}-\frac{q^2}{m^2}\right)\arctan\left(\frac qm\right)\right].
\label{Gclos}
\ee
In the deep IR ($\qe\to0$), and for asymptotically large momenta ($\qe\to\infty$), one finds
\be
G(\qe)\stackrel{\qe\to0}{\longrightarrow}-\frac{g^2 C_A}{6\pi m},
\qquad\qquad
G(\qe)\stackrel{\qe\to\infty}{\longrightarrow}0.
\label{Glim}
\ee

From the expressions for $\widehat\Pi_m(q)$ and $G(q)$ obtained above, we can 
determine the conventional gluon propagator, $\Delta(\qe)$, 
in the Landau gauge; the latter can then be compared to the lattice data. To that end, let us first employ 
the crucial identity of (\ref{bqi2}) to write 
\be
\Delta(\qe)=\frac{[1+G(\qe)]^2}{\qe^2+m^2 + \widehat{\Pi}_m(\qe)}.
\ee
Then, by virtue of (\ref{Glim}), $\Delta(\qe)$ has the same asymptotic behavior as 
$\widehat{\Delta}(\qe)$. 

\begin{figure}[!t]
\begin{center}
\includegraphics{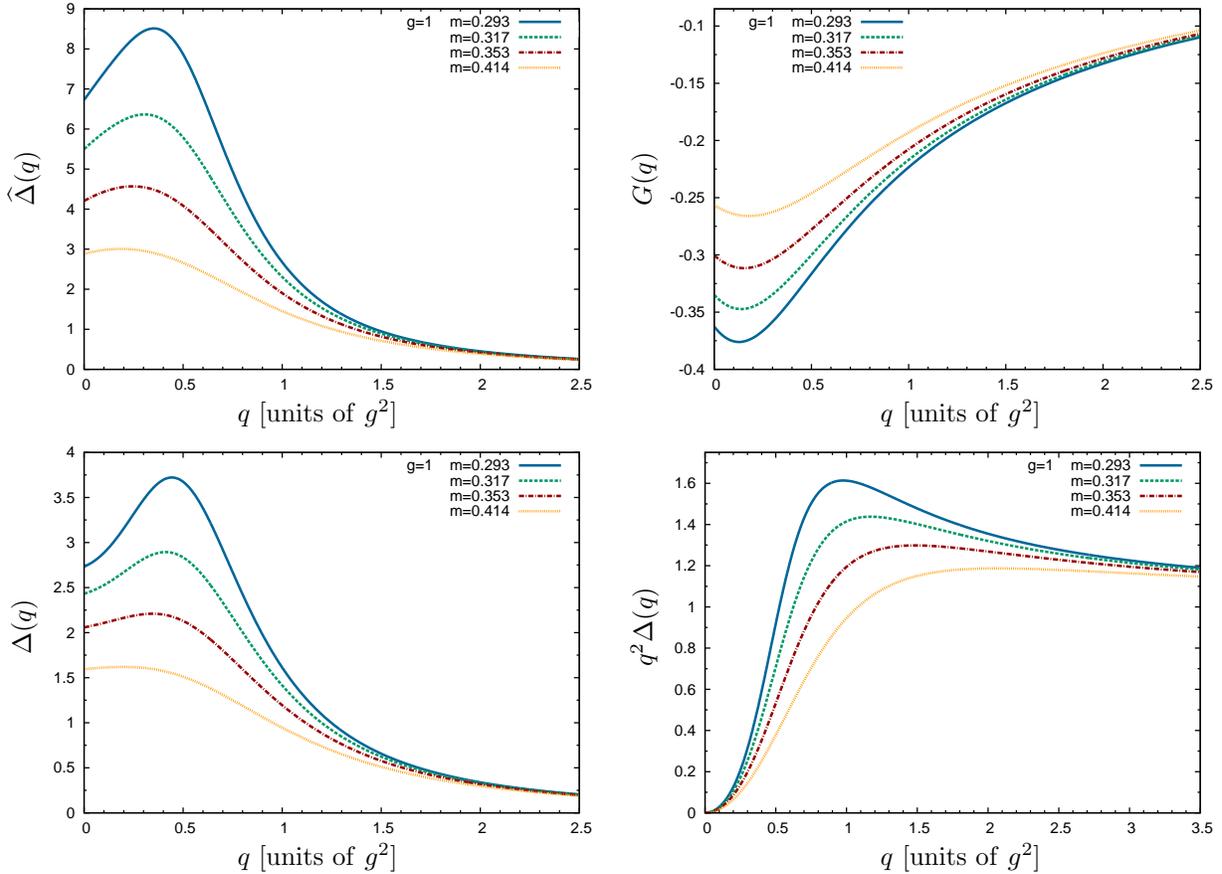}
\end{center}
\caption{\label{analytic-figure-1}Results for the massive one-loop approximation for the $d=3$ gluon propagator. In the upper panels we show the plots for different values of the hard-mass parameter $m$ for the background-quantity identity ingredients $\widehat{\Delta}(q)$ (left) and the Kugo-Ojima function $G(q)$ (right). In the lower panels we show the conventional propagator $\Delta(q)$ (left) and its corresponding dressing function $q^2\Delta(q)$ (right).}
\end{figure}

Notice that in $d=3$ the gluon  and ghost propagators have the basic scaling property
\be
\Delta(q,g,m)=a^{2}\Delta(aq,\sqrt{a}g,am), \qquad D(q,g,m)=a^{2}D(aq,\sqrt{a}g,am), 
\label{scaling}
\ee
where $a$ is a positive real number. 
Of course, the corresponding dressing functions 
(being dimensionless quantities) are invariant under such a combined rescaling;
for example, the ghost dressing function satisfies $F(q,g,m)=F(aq,\sqrt{a}g,am)$, and so does 
the gluon dressing function $q^2\Delta(q)$ and the Kugo-Ojima function $G(q)$.
One can then make use of these scaling properties to set $g$ (respectively $m$) equal to unity, and vary $m$ (respectively $g$) 
in order to study the shape of the solutions found so far. The results (when setting $g=1$ and varying $m$) are shown in Fig.~\ref{analytic-figure-1}.

\subsection{Comparing the gluon propagator with $SU(2)$ lattice data}

\begin{figure}[!t]
\begin{center}
\includegraphics{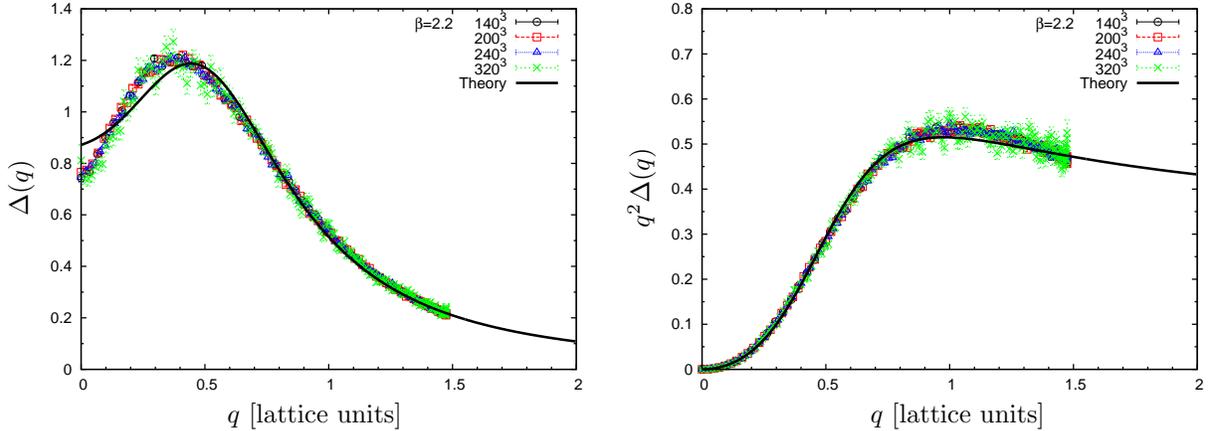}
\end{center}
\caption{\label{bestfit-figure}Comparison of the lattice results of~\cite{Cucchieri:2003di} with the gluon propagator (left) and the gluon dressing function (right) obtained within the massive one-loop approximation adopted in this paper. In passing, notice that the dressing function does not tend to 1 for asymptotically large $q$ which also motivates the momentum rescaling procedure employed.} 
\end{figure}

We next compare the result of our calculation for the conventional gluon propagator $\Delta(q)$
with the lattice results of~\cite{Cucchieri:2003di}. In order to do that, 
the lattice data must be first properly normalized (or, equivalently, 
the theoretical prediction must be suitably rescaled)
Specifically, 
in the absence of any physical input that would fix 
the physical scale, one uses the scaling property (\ref{scaling})
and determine the scaling factor $a$ in such a way that the 
asymptotic (large momentum) segment of the lattice data coincides with that  
obtained from our calculation; indeed the two ``tails'' should coincide, given that perturbation theory is reliable
in that region of momenta. 
The result of this procedure is shown in Fig.~\ref{bestfit-figure}; evidently, the 
matching between the theoretical curve and the lattice data is very good. 
The best-fit curve furnishes the ratio
\be
\frac{m}{2 g^2}\approx0.146,
\ee
which appears to be in rather good agreement with previous theoretical and lattice studies~\cite{Alexanian:1995rp}.

\subsection{Ghost dressing function and lattice data}

\begin{figure}[!t]
\begin{center}
\includegraphics{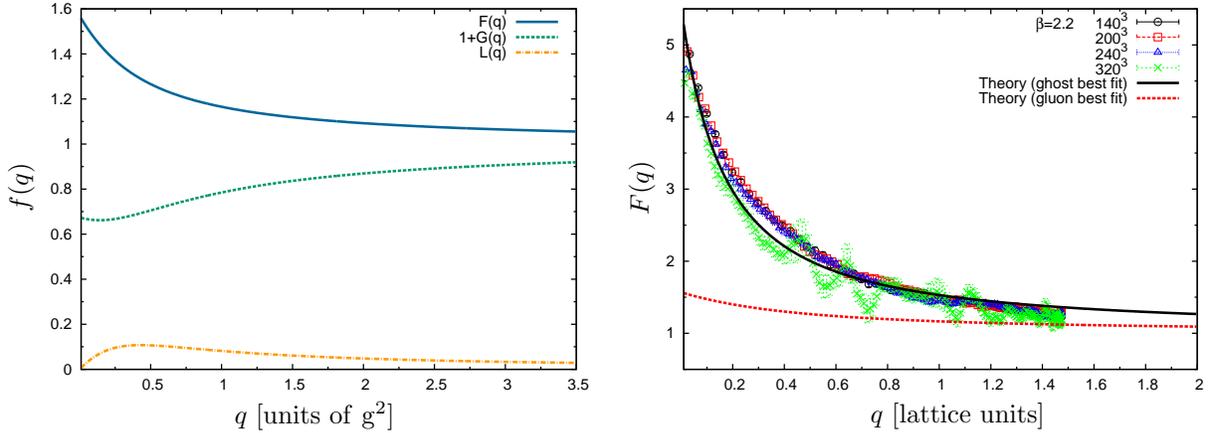}
\end{center}
\caption{\label{ghost-figure}(Left panel) Values of $L(q)$, $1+G(q)$,  and $F(q)=[1+G(q)+L(q)]^{-1}$ within our approximation for $g=1.287$ and $m=0.539$. (Right panel) Comparison of the ghost dressing function with the one calculated within our approximation for two sets of values corresponding to the gluon propagator best fit ($g=1.287$ and $m=0.539$, red dashed line) and to the best fit to the lattice data ($g=2.049$, $m=0.543$).} 
\end{figure}

We next proceed to calculate the theoretical prediction for the ghost-dressing function 
$F(q)$. In a spirit similar to that adopted for the gluon propagator,  
as first approach in this direction, we simply compute the  
diagram for the ghost propagator (see Fig.~\ref{ghostSDE}) 
using as inputs on the rhs  $\Delta \to \Delta_m$ and $D \to D_0 $.
The result of this calculation is
\bea
F(q)&=& 1+i g^2 C_A \int_k\frac1{(k+q)^2 (k^2-m^2)}\left[1-\frac{(k\cdot q)^2}{k^2q^2}\right]
\nonumber\\
&=&1+ \frac{ig^2C_A}4\left[\frac2{q^2}I_4+3I_2+I_3-
\frac{q^2}{m^2}\left(I_2-I_1 \right)\right]
\label{Fint}
\eea
At this point, and before attempting a comparison with the corresponding lattice data, 
we note that, in the Landau gauge only, 
the ghost-dressing function $F(q)$, and the two form factors $G(q)$ and $L(q)$ 
defined in Eq.~(\ref{LDec}), are related (for any $d$) by the following 
important identity, 
\be
1+ G(q) + L(q) = F^{-1}(q).
\label{funrel}
\ee
The relation of Eq.~(\ref{funrel}), 
has been first obtained in~\cite{Kugo:1995km}, and some years later in~\cite{Grassi:2004yq}, 
in the framework of the Batalin-Vilkovisky quantization formalism; 
as was shown there, this relation is a direct consequence of the fundamental BRST symmetry.
Recently, the same identity has been derived exactly from the SDEs of the theory~\cite{Aguilar:2009nf}, 
and the important property $L(0)=0$, usually assumed in the literature, 
was shown to be valid for any value of the space-time dimension $d$; indeed 
setting  $\Delta \to \Delta_m$ and $D \to D_0 $ on the rhs of Eq.~(\ref{Lapp}), one has
\bea
L(q)&=&\frac{ig^2C_{\chic A}}2\int_k\!
\frac1{(k+q)^2(k^2-m^2)}\left[1-3\frac{(k\cdot q)^2}{k^2q^2}\right]\nonumber \\
&=&\frac{ig^2C_{\chic A}}8\left[\frac6{q^2}I_4+3I_3+I_2-3\frac{q^2}{m^2}(I_2-I_1)\right]\nonumber \\
&=& \frac{3g^2 C_A}{32\pi m}\left[\frac\pi2\frac{q}{m}+\frac{m^2}{q^2}-1+\left(\frac23-\frac{m^2}{q^2}-\frac{q^2}{m^2}\right)\arctan\left(\frac qm\right)\right],
\label{Lclos}
\eea
and therefore
\be
L(\qe)\stackrel{\qe\to0}{\longrightarrow}0,
\qquad\qquad
L(\qe)\stackrel{\qe\to\infty}{\longrightarrow}0.
\label{Llim}
\ee
It is then straightforward to verify from the result above and the closed 
expressions given in Eqs~(\ref{Gclos}) and~(\ref{Fint}), that Eq.~(\ref{funrel}) holds exactly within the 
approximation scheme we are using (see also the left panel of Fig.~\ref{ghost-figure}). 

We next vary the parameters $g$ and $m$ in the expression given in Eq.~(\ref{Fint}) 
in order to reproduce the lattice data for $F(q)$. As a natural starting point 
we use the values that have resulted in the best fit for the gluon propagator, 
namely $g=1.285$ and $m=0.480$. However, as is clear from the red dashed curve shown in~Fig.~\ref{ghost-figure} (right panel), 
the result obtained is in poor agreement with the lattice.
If instead we allow the parameters to vary freely, {\it i.e.}, we disregard the 
gluon data and attempt to only reproduce the ghost data,
the best possible curve is shown by the black continuous line of~Fig.~\ref{ghost-figure}, being obtained 
for the values $g=2.049$, $m=0.543$ (giving $m/2g^2\approx0.065$). 

It is clear from this analysis that, within the approximation scheme employed, 
the lattice data may be well reproduced if treated independently, 
but it is not possible to arrive at a  reasonable simultaneous fit, {\it i.e.}, to fit 
both curves using a unique set of parameters.

\subsection{Combined treatment: gluon propagator and ghost dressing function vs lattice}

To remedy this situation, we will improve the approximation used for 
obtaining the theoretical prediction for the ghost-dressing function.  
Specifically,  
we will study an approximate version of the ghost SDE given in Eq.~(\ref{SDgh}), and we will 
{\it solve} self-consistently for the unknown function $F(q)$, 
instead of simply calculating its rhs, as was done above for obtaining the expression in Eq.~(\ref{Fint}).

Given that Eq.~(\ref{SDgh}) contains $\Delta (k)$ as one of its basic ingredient, 
the general matching procedure becomes more subtle.
In particular, instead of freely fitting just one set of data (that for the gluon propagator) 
one must now attempt to fit simultaneously both the gluon and ghost data, as well as possible.  
As we will see, this more complicated procedure furnishes finally 
a very good agreement with the combined set of lattice data, but 
one has to settle for a slightly less 
accurate description of the gluon data compared to the one obtained in Fig.~\ref{bestfit-figure}

After approximating the gluon-ghost vertex $\Gamma_{\mu}$ by its tree-level value, 
we arrive at the following integral equation for the ghost dressing function $F$,
\be
F^{-1}(q) = 1 +g^2 C_A \int_k \left[1 - \frac{(k \cdot q)^2}{k^2 q^2}\right] \frac{\Delta (k)F(k+q)}{(k+q)^2}\,.
\label{tt2}
\ee 
The general idea now is to solve Eq.~(\ref{tt2}) numerically for $F(q)$, 
using as input for the $\Delta (k)$ under the integral sign 
the theoretical curve that,  
after the rescaling mentioned earlier, provides the best possible fit to the gluon data, and, at the same time,
allows for the numerical convergence of Eq.~(\ref{tt2}).  
We note in passing that this procedure permits, 
after a shift of the integration variable, 
to pass all angular dependence from $F(k+q)$ to $\Delta (k)$, whose functional form  
is considered known;  as a result, one does not need to resort to further approximations 
for the angular part of the integral equation. 

The general observation regarding the 
numerical treatment of Eq.~(\ref{tt2}) is that it appears to  
be extremely sensitive to the precise shape of $\Delta (k)$ and the value of $g$; 
minute variations of these quantities give rise to large disparities in the resulting  $F(q)$.

\begin{figure}[!t]
\begin{center}
\includegraphics{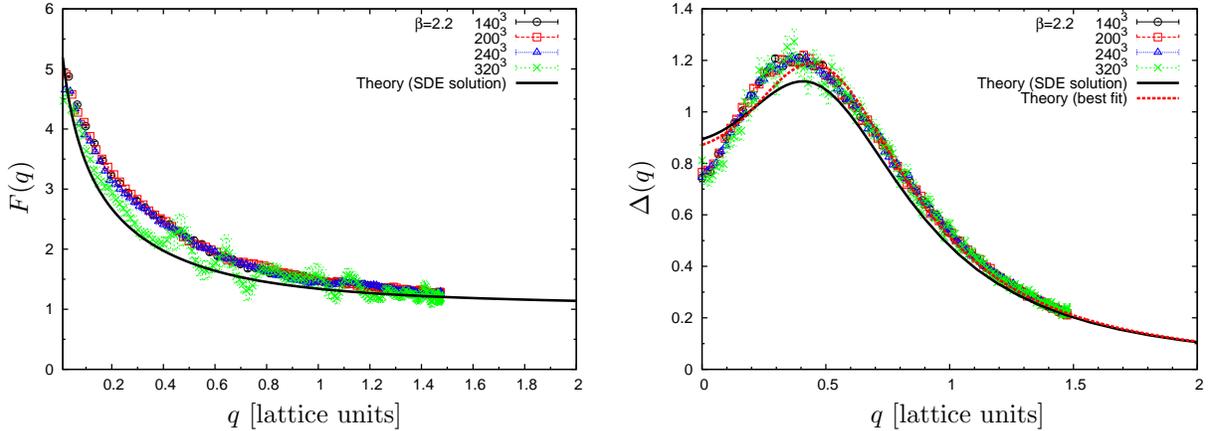}
\end{center}
\caption{\label{gluon-ghost-figure}Comparison of the lattice results of~\cite{Cucchieri:2003di} with the ghost dressing function obtained through the solution of the ghost SDE (left); on the right we show the gluon propagator for the same value of $m/2g^2\approx0.153$.} 
\end{figure}

The best possible solution that we have obtained is shown in the left panel of Fig.~\ref{gluon-ghost-figure} 
As announced, the accuracy achieved in matching the lattice data 
for the gluon propagator is slightly inferior to that of our best fit (Fig.~\ref{gluon-ghost-figure}, right panel), but is still very good. 

\section{Conclusions}

In this work we have presented a nonperturbative study of the 
(Landau gauge) gluon and ghost propagator for $d=3$ Yang-Mills, 
using the ``one-loop dressed'' SDEs of the PT-BFM formalism. 
One of the most powerful features of this framework is that the 
transversality of the truncated gluon self-energy is guaranteed, 
by virtue of the QED-like Ward identities satisfied by the fully-dressed vertices 
entering into the dynamical equations. 

The central dynamical ingredient of our analysis is
the assumption that the famous Schwinger mechanism,  
namely the dynamical formation of zero-mass
Nambu-Golstone-boson-like composite excitations, which 
allow the gauge-invariant generation of a gauge-boson mass, 
is indeed realized in $d=3$ Yang-Mills.
The way this dynamical scenario is  incorporated into the SDEs 
is through the form of the three-gluon vertex. 
Specifically, in order to satisfy the correct Ward identity, 
as required by gauge-invariance, this vertex must contain 
massless, longitudinally coupled poles, representative 
precisely of the aforementioned composite excitations.

It should be emphasized that the approach followed here 
is approximate, not only in the sense that we consider the 
one-loop dressed version of the SDE, omitting (gauge-invariantly) higher orders 
[{\it i.e.}, the third and fourth block of Fig.~\ref{sde_bfm}], 
but also because 
we do not actually solve simultaneously the full system of resulting equations. 
Specifically, as explained in Section IV, when evaluating the gluon self-energy 
we have used tree-level expressions for the ghost propagators appearing in 
diagram $(a_3)$ of Fig.~\ref{sde_bfm}, and the same approximation 
is used also in the determination of the KO function $G(q)$. 
We have then used the resulting gluon propagator 
as an input into Eq.~(\ref{tt2}) to obtain the improved $F(q)$. 
Of course, this two-step procedure is bound to result in 
a considerable discrepancy between the ``one-loop'' $G(q)$ and the 
$F(q)$ obtained from solving its corresponding SDE; evidently, the identity 
of Eq.~(\ref{funrel}) cannot be fulfilled any longer.
In addition, the dynamical gluon mass $m$ has been treated for simplicity as a constant.
However, a more thorough study should eventually include the important feature  
that the mass depends nontrivially on the momentum, in accordance with general considerations~\cite{Lavelle:1991ve};
in fact, a complete 
SDE treatment ought to actually {\it determine} the precise way the mass is running~\cite{Aguilar:2009ke}.
The fact that, despite these simplifications, the lattice results 
for the gluon and ghost propagator are so well reproduced, 
suggests that the full treatment may reveal a number of subtle   
cancellations, caused by the highly non-linear nature of the SDE equations,  
yielding finally results very similar to those reported here.

Let us finally outline briefly some of the modifications and additional 
field-theoretic inputs that such a full SDE treatment would entail.
To begin with, 
a more complete Ansatz for the three-gluon vertex ${\g}_{\alpha\mu\nu}$, appearing in the SDE 
of the gluon propagator [graph $(a_1)$ in Fig.~\ref{sde_bfm}], 
must be devised. Such an Ansatz must not contain only the 
part of the massless poles, as Eq.~(\ref{Ver1}) does, which only accounts for the massive part of the 
propagator, 
but should make explicit reference to the 
entire $\Delta$, in the spirit of the analysis already presented in~\cite{Aguilar:2009ke}
(for $d=4$). In addition, a similar Ansatz must be 
introduced also for the full gluon-ghost vertex ${\g}_{\alpha}$ appearing in 
the graph $(a_3)$ of Fig.~\ref{sde_bfm}. In order to maintain explicit gauge-invariance, ${\g}_{\alpha}$ must be 
such that the second Ward identity of Eq.~(\ref{fourWI}) is automatically satisfied.  
Note that ${\g}_{\alpha}$ is not the same as the conventional gluon-ghost vertex 
$\gb_{\alpha}$ that appears in Eq.~(\ref{SDgh}) and in Fig.~\ref{ghostSDE}. 
Given that ${\g}_{\alpha}$ and $\gb_{\alpha}$ are different, and that only the former  
is crucial for the transversality of the gluon SDE, one may 
approximate  $\gb_{\alpha}$ by its tree-level value, 
without clashing with gauge-invariance, a freedom that exists only 
within the PT-BFM scheme. Should one opt for a more sophisticated treatment of $\gb_{\alpha}$, 
then an appropriate Ansatz may have to be devised.  
Given that $\gb_{\alpha}$ satisfies a complicated Slavnov-Taylor identity, instead of the 
simple Ward identity of ${\g}_{\alpha}$ , further approximations may be necessary. 
We hope to be able to implement some of the aforementioned improvements in the near future.  

\acknowledgments 
\vspace{-0.5cm}
We would like to thank  A. Cucchieri and T. Mendes 
for kindly making their lattice results available to us, and for their useful comments.  
The research of J.~P. is supported by the European FEDER and  Spanish MICINN under grant FPA2008-02878, and the Fundaci\'on General of the UV. The work of  A.C.A  is supported by the Brazilian Funding Agency CNPq under the grant 305850/2009-1.


\vspace{-0.5cm}


\begin{thebibliography}{99}


\bibitem{Gross:1980br}
  D.~J.~Gross, R.~D.~Pisarski and L.~G.~Yaffe,
  Rev.\ Mod.\ Phys.\  {\bf 53}, 43 (1981).


\bibitem{Jackiw:1980kv}
  R.~Jackiw and S.~Templeton,
  Phys.\ Rev.\  D {\bf 23}, 2291 (1981).


\bibitem{Appelquist:1981vg}
  T.~Appelquist and R.~D.~Pisarski,
  Phys.\ Rev.\  D {\bf 23}, 2305 (1981).

\bibitem{Deser:1981wh}
  S.~Deser, R.~Jackiw and S.~Templeton,
  Annals Phys.\  {\bf 140}, 372 (1982)


\bibitem{Deser:1982vy}
  S.~Deser, R.~Jackiw and S.~Templeton,
  Phys.\ Rev.\ Lett.\  {\bf 48}, 975 (1982).


\bibitem{Nadkarni:1982kb}
  S.~Nadkarni,
  Phys.\ Rev.\  D {\bf 27}, 917 (1983).

\bibitem{Cornwall:1984eu}
  J.~M.~Cornwall, W.~S.~Hou and J.~E.~King,
  Phys.\ Lett.\  B {\bf 153}, 173 (1985).


\bibitem{Nadkarni:1988ti}
  S.~Nadkarni,
  Phys.\ Rev.\ Lett.\  {\bf 61}, 396 (1988).


\bibitem{Cornwall:1988ad}
  J.~M.~Cornwall,
  Physica A {\bf 158}, 97 (1989).


\bibitem{Kobes:1990dc}
  R.~Kobes, G.~Kunstatter and A.~Rebhan,
  Nucl.\ Phys.\  B {\bf 355}, 1 (1991).


\bibitem{Cornwall:1992cu}
  J.~M.~Cornwall,
  Nucl.\ Phys.\  B {\bf 416}, 335 (1994).

\bibitem{Buchmuller:1994qy}
  W.~Buchmuller and O.~Philipsen,
  Nucl.\ Phys.\  B {\bf 443}, 47 (1995).


\bibitem{Jackiw:1995nf}
  R.~Jackiw and S.~Y.~Pi,
  Phys.\ Lett.\  B {\bf 368}, 131 (1996).


\bibitem{Alexanian:1995rp}
  G.~Alexanian and V.~P.~Nair,
  Phys.\ Lett.\  B {\bf 352}, 435 (1995).


\bibitem{Cornwall:1995ac}
  J.~M.~Cornwall and B.~Yan,
  Phys.\ Rev.\  D {\bf 53}, 4638 (1996)

\bibitem{Cornwall:1996jb}
  J.~M.~Cornwall,
  Phys.\ Rev.\  D {\bf 54}, 1814 (1996)



\bibitem{Karsch:1996aw}
  F.~Karsch, T.~Neuhaus, A.~Patkos and J.~Rank,
  Nucl.\ Phys.\  B {\bf 474}, 217 (1996).


\bibitem{Jackiw:1997jga}
  R.~Jackiw and S.~Y.~Pi,
  Phys.\ Lett.\  B {\bf 403}, 297 (1997).

\bibitem{Buchmuller:1996pp}
  W.~Buchmuller and O.~Philipsen,
  Phys.\ Lett.\  B {\bf 397}, 112 (1997).

  
\bibitem{Cornwall:1997dc}
  J.~M.~Cornwall,
  Phys.\ Rev.\  D {\bf 57}, 3694 (1998).


\bibitem{Karabali:1998yq}
  D.~Karabali, C.~j.~Kim and V.~P.~Nair,
  Phys.\ Lett.\  B {\bf 434}, 103 (1998).

  
\bibitem{Eberlein:1998yk}
F.~Eberlein,
Phys.\ Lett.\  B {\bf 439}, 130 (1998).


\bibitem{Heller:1997nqa}
  U.~M.~Heller, F.~Karsch and J.~Rank,
  Phys.\ Rev.\  D {\bf 57}, 1438 (1998).

\bibitem{Eberlein:1998mb}
 F.~Eberlein,
 Nucl.\ Phys.\  B {\bf 550}, 303 (1999).



\bibitem{Cucchieri:2001tw}
  A.~Cucchieri, F.~Karsch and P.~Petreczky,
  Phys.\ Rev.\  D {\bf 64}, 036001 (2001)

\bibitem{Nakamura:2003pu}
  A.~Nakamura, T.~Saito and S.~Sakai,
  Phys.\ Rev.\  D {\bf 69}, 014506 (2004)


\bibitem{Burgio:2009xp}
  G.~Burgio, M.~Quandt and H.~Reinhardt,
  arXiv:0911.5101 [hep-lat].

\bibitem{Quandt:2010yq}
  M.~Quandt, H.~Reinhardt and G.~Burgio,
  arXiv:1001.3699 [hep-lat].





\bibitem{Cornwall:1982zr}
J.~M.~Cornwall,
Phys.\ Rev.\ D {\bf 26}, 1453 (1982). 



\bibitem{Schwinger:1962tn}
  J.~S.~Schwinger,
  Phys.\ Rev.\  {\bf 125}, 397 (1962).

\bibitem{Schwinger:1962tp}
  J.~S.~Schwinger,
  Phys.\ Rev.\  {\bf 128}, 2425 (1962).

\bibitem{Jackiw:1973tr}
  R.~Jackiw and K.~Johnson,
  Phys.\ Rev.\ D {\bf 8}, 2386 (1973).

\bibitem{Cornwall:1973ts}
  J.~M.~Cornwall and R.~E.~Norton,
  Phys.\ Rev.\ D {\bf 8} 3338 (1973).

\bibitem{Eichten:1974et}
E.~Eichten and F.~Feinberg,
Phys.\ Rev.\ D {\bf 10}, 3254 (1974).

\bibitem{Jackiw:1973ha}
  R.~Jackiw,
  ``Dynamical Symmetry Breaking,''
{\it  In *Erice 1973, Proceedings, Laws Of Hadronic Matter*, New York 1975, 225-251 and M I T Cambridge - COO-3069-190 (73,REC.AUG 74) 23p}.

\bibitem{Farhi:1982vt}
  E.~Farhi and R.~Jackiw,
  ``Dynamical Gauge Symmetry Breaking. A Collection Of Reprints,''
{\it  Singapore, Singapore: World Scientific ( 1982) 403p}.


\bibitem{Frampton:2008zz}
  P.~H.~Frampton,
  ``Gauge Field Theories: Third Revised and Improved Edition,''
{\it  John Wiley and Sons, Inc, 2008}.

\bibitem{Aguilar:2008xm}
  A.~C.~Aguilar, D.~Binosi and J.~Papavassiliou,
  Phys.\ Rev.\  D {\bf 78}, 025010 (2008).


\bibitem{Gribov:1977wm}
 V.~N.~Gribov,
  Nucl.\ Phys.\  B {\bf 139}, 1 (1978).

\bibitem{Zwanziger:1993dh}
  D.~Zwanziger,
  Nucl.\ Phys.\  B {\bf 412}, 657 (1994).

\bibitem{Dudal:2008sp}
  D.~Dudal, J.~A.~Gracey, S.~P.~Sorella, N.~Vandersickel and H.~Verschelde,
  Phys.\ Rev.\  D {\bf 78}, 065047 (2008);
  Phys.\ Rev.\  D {\bf 78}, 125012 (2008)


\bibitem{Cornwall:1989gv}
  J.~M.~Cornwall and J.~Papavassiliou,
  Phys.\ Rev.\  D {\bf 40}, 3474 (1989).

\bibitem{Binosi:2002ft}
  D.~Binosi and J.~Papavassiliou,
  Phys.\ Rev.\  D {\bf 66}(R), 111901 (2002);
 D.~Binosi and J.~Papavassiliou,
  J.\ Phys.\ G {\bf 30}, 203 (2004).
  
\bibitem{Binosi:2009qm}
  D.~Binosi and J.~Papavassiliou,
  Phys.\ Rept.\  {\bf 479}, 1 (2009).

\bibitem{Nair:2005iw}
 V.~P.~Nair,
``Quantum field theory: A modern perspective,''
{\it  New York, USA: Springer (2005) 557 p}.


\bibitem{Abbott:1980hw}
L.~F.~Abbott,
Nucl.\ Phys.\ B {\bf 185}, 189 (1981).

\bibitem{Aguilar:2006gr}
  A.~C.~Aguilar and J.~Papavassiliou,
  JHEP {\bf 0612}, 012 (2006)

\bibitem{Binosi:2007pi}
  D.~Binosi and J.~Papavassiliou,
  Phys.\ Rev.\  D {\bf 77}, 061702 (2008);
  JHEP {\bf 0811}, 063 (2008).
  

\bibitem{Cucchieri:2003di}
  A.~Cucchieri, T.~Mendes and A.~R.~Taurines,
  Phys.\ Rev.\  D {\bf 67}, 091502 (2003);
  A.~Cucchieri and T.~Mendes,
 PoS(QCD-TNT09)026 (2010).


\bibitem{Kugo:1979gm}
  T.~Kugo and I.~Ojima,
  Prog.\ Theor.\ Phys.\ Suppl.\  {\bf 66}, 1 (1979).

\bibitem{Grassi:1999tp}
  P.~A.~Grassi, T.~Hurth and M.~Steinhauser,
  Annals Phys.\  {\bf 288}, 197 (2001);
  D.~Binosi and J.~Papavassiliou,
  Phys.\ Rev.\ D {\bf 66}, 025024 (2002).






\bibitem{Poggio:1974qs}
 E.~C.~Poggio, E.~Tomboulis and S.~H.~Tye,
 Phys.\ Rev.\  D {\bf 11}, 2839 (1975)).



\bibitem{Kugo:1995km}
  T.~Kugo,
  arXiv:hep-th/9511033.


\bibitem{Grassi:2004yq}
  P.~A.~Grassi, T.~Hurth and A.~Quadri,
  Phys.\ Rev.\  D {\bf 70}, 105014 (2004).


\bibitem{Aguilar:2009nf}
  A.~C.~Aguilar, D.~Binosi, J.~Papavassiliou and J.~Rodriguez-Quintero,
  Phys.\ Rev.\  D {\bf 80}, 085018 (2009)

\bibitem{Lavelle:1991ve}
  M.~Lavelle,
  Phys.\ Rev.\  D {\bf 44}, 26 (1991).

\bibitem{Aguilar:2009ke}
  A.~C.~Aguilar and J.~Papavassiliou,
  Phys.\ Rev.\  D {\bf 81}, 034003 (2010).


\end{thebibliography}
\end{document}